# On the Absence of Triplet Exciton Loss Pathways in Non-Fullerene Acceptor based Organic Solar Cells

Maria S. Kotova[1], Giacomo Londi[2], Johannes Junker[1], Stefanie Dietz[1], Alberto Privitera[3], Kristofer Tvingstedt[1], David Beljonne[2], Andreas Sperlich[1*] and Vladimir Dyakonov[1]

[1] Experimental Physics 6, Julius Maximilian University of Würzburg, Am Hubland, 97074 Würzburg, Germany

[2] Laboratory for Chemistry of Novel Materials, University of Mons, B-7000 Mons, Belgium

[3] Clarendon Laboratory, Department of Physics, University of Oxford, Oxford OX1 3PU, England, UK

Abstract

We investigate the viability of highly efficient organic solar cells (OSCs) based on non-fullerene acceptors (NFA) by taking into consideration efficiency loss channels and stability issues caused by triplet excitons (TE) formation. OSCs based on a blend of the conjugated donor polymer PBDB-T and ITIC as acceptor were fabricated and investigated with electrical, optical and spin-sensitive methods. The spin-Hamiltonian parameters of molecular TEs and charge transfer TEs in ITIC e.g., zero-field splitting and charge distribution, were calculated by Density Functional Theory (DFT) modelling. In addition, the energetic model describing the photophysical processes in the donor-acceptor blend was derived. Spin-sensitive photoluminescence measurements prove the formation of charge transfer (CT) states in the blend and the formation of TEs in the pure materials and the blend. However, no molecular TE signal is observed in the completed devices under working conditions by spin-sensitive electrical measurements. The absence of a molecular triplet state population allows to eliminate a charge carrier loss channel and irreversible photooxidation facilitated by long-lived triplet states. These results correlate well with the high power conversion efficiency of the PBDB-T:ITIC-based OSCs and their high stability.

1. Introduction

A new era of organic photovoltaic (OPV) research started after the introduction of new acceptor (A) and donor (D) materials in the technology. In the past, most of the bulk heterojunction (BHJ) organic solar cells (OSCs) employed soluble fullerene derivatives as electron acceptors due to their superior electron affinity and good transport properties. Nevertheless, fullerene acceptors have limited absorption of the solar spectrum, air- and light-induced degradation issues and it is difficult to modify their energy levels. Development of novel non-fullerene acceptors (NFA) and good matching donor polymers, that overcome some of these issues[1], have led to rapid progress and power conversion efficiencies (PCE) of over 18% have been achieved in single junction devices [2,3,4]. This success has been achieved in part by the application of thiophene based NFAs, e.g., 3,9-bis(2-methylene-(3-(1,1-dicyanomethylene)-indanone))-5,5,11,11-tetrakis(4-hexylphenyl)-dithieno[2,3-d:2′,3′-d′]-s-indaceno[1,2-b:5,6-b′]dithiophene (ITIC) or its derivatives. ITIC is a fused ring electron acceptor[5], consisting of donor core and strong electron acceptor units at the two sides of the backbone. Despite this rapid rise in efficiency, a fundamental understanding of the charge generation mechanism and loss channels is still lacking.

*sperlich@physik.uni-wuerzburg.de

Upon photoexcitation, bound electron-hole pairs, excitons, are generated in D and A materials. Excitons dissociate at the D/A interface and form interfacial charge transfer (CT) states. It has been widely reported that free charges can be generated from these intermediate CT states. The energy of the CT state ($E_{CT}$) is closely related to the energetic difference between the donor HOMO and the acceptor LUMO. This value represents a fundamental limit of the open-circuit voltage ($V_{OC}$). The remarkably high $V_{OC}$ in NFA-based OSCs has been achieved by increasing $E_{CT}$ via a closer matching of A and D energy levels. However, an increase in $E_{CT}$ can lead to a new loss pathway resulting from charge recombination to the now energetically favorable D and A triplet exciton (TE) states [6, 7, 8, 9, 10, 11]. Formation of TEs is not only causing a new loss pathway and reduced short-circuit current ($J_{SC}$) but can also lead to enhanced degradation of the active layer. The energy of TEs is sufficient to excite ground state triplet oxygen ($^3O_2$) adsorbed from ambient air to its very reactive excited singlet form ($^1O_2^*$) [12]. This can result in chemical reactions of the OSC's active layer with the excited singlet oxygen and finally in a degradation of OSC performance[13].

The presence of the TE population in NFA-based OSCs and their involvement in carrier leakage is investigated in this work using two spectroscopic techniques: photoluminescence detected magnetic resonance (PLDMR) on thin active layer films and electrically detected magnetic resonance (EDMR) on fully processed OSCs under operating conditions. To pinpoint energy levels of all involved excited states (singlet/triplet excitons, CT states) additional photoluminescence (PL), external quantum efficiency (EQE), temperature dependent current density-voltage dependence ($J$-$V$) and electroluminescence (EL) were studied on pure materials, PBDB-T:ITIC blends and solar cells based on them. In order to support spin-sensitive measurements, DFT modelling was performed as well. The analysis showed that the population of donor and acceptor triplet states does take place in pure donor and acceptor materials and in their blends at low temperatures. Nevertheless, charge separation and extraction outperform the less efficient triplet formation and recombination in working devices at ambient conditions.

2. Experimental Section

    2.1 Materials and Devices

ITIC was used as acceptor and poly[(2,6-(4,8-bis(5-(2-ethylhexyl)thiophen-2-yl)benzo[1,2-b:4,5-b′]dithiophene)-co-(1,3-di(5-thiophene-2-yl)-5,7-bis(2-ethylhexyl)-benzo[1,2-c:4,5-c′]dithiophene-4,8-dione))] (PBDB-T) as donor for the BHJ active layer. PBDB-T is known for good thermal stability and has good HOMO alignment with ITIC. This material combination was shown to be very efficient in solar cells with a maximum reported PCE of 11.3% [14].

ITIC and PBDB-T were purchased from 1-Material, PEDOT:PSS from Heraeus, Al doped ZnO (AZO) nanoparticle solution from Avantama (slot N-21X) and all were used without further purification. Sample preparation was done on Herasil ITO-glass substrates.

Devices were fabricated in the conventional ITO/PEDOT:PSS/active layer/Ca/Al and inverted ITO/ZnO/active layer/MoO3/Al device structures (Figure 1). The ITO-coated glass substrates were thoroughly cleaned by deionized water, acetone, isopropanol and etched in oxygen plasma for 30 s. The cleaned substrates were then covered with a thin layer (25 nm) of PEDOT:PSS or AZO. PEDOT:PSS solution was used without modifications, coated substrates were annealed at 130°C for 15 minutes in air. AZO solution was diluted with isopropanol at 1:1 concentration. Coated substrates were then annealed at 120°C for 10 minutes in air. Further processing steps were performed inside a nitrogen glovebox. PBDB-T and ITIC were dissolved in chlorobenzene solvent with a total concentration of 20 mg/mL in ratio of PBDB-T:ITIC 1:1 and stirred at 50˚C for at least 24 hours. Subsequently, the mixture was spin coated on AZO or PEDOT:PSS covered glass substrates to reach an optimal film thickness of approximately 100 nm. The active layers were thermally annealed for 30 minutes at 70°C. Finally, 10 nm $MoO_3$ and 100 nm Ag or alternatively 5 nm Ca and 120 nm Al layers were deposited subsequently to complete the inverted or conventional devices, respectively. Devices were fabricated in various geometries with active areas of 1.5 $mm^2$, 3 $mm^2$ and 9 $mm^2$.

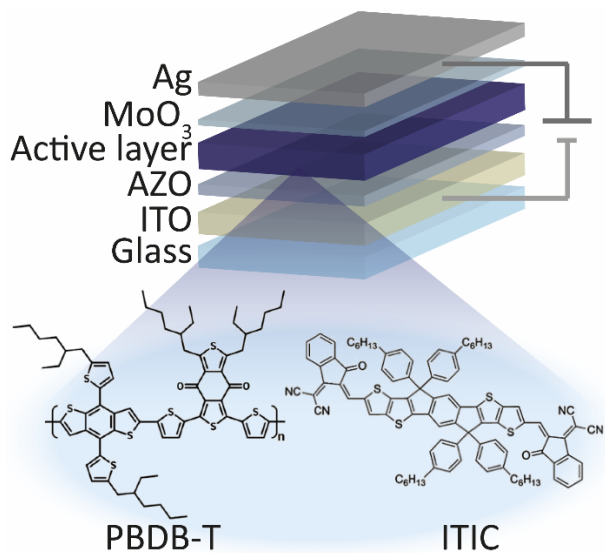

**Figure 1.** Structure of inverted solar cells.

Samples for photoluminescence detected magnetic resonance (PLDMR) measurements were prepared in EPR glass tubes. For this, about 100 µl solution of the investigated material was filled into a standard X-band EPR tube with 3 mm inner diameter. Drying was achieved via evacuation to a rough vacuum of $10^{-2}$ mbar and flushing at least three times with helium gas. During the drying process, a thin layer of the investigated material is deposited on the inner walls of the tube. After the drying procedure, the tubes were sealed with a blow torch.

2.2 Methods

The current density-voltage (*J-V*) measurements were carried out in a $N_2$ glove box at room temperature under 100 mW $cm^{-2}$ illumination of the standard AM 1.5G spectrum. *J-V* curves were measured with a Keithley 2612B programmable current–voltage source. The thicknesses of films were measured with a Dektak profilometer.
Photoluminescence (PL) is provided by exciting the sample from a side with a 405 nm, 4.5 mW CPS405 cw laser diode module from Thorlabs. Electroluminescence (EL) is generated by current injection by an Agilent 4155C parameter analyzer for 10 or 20 seconds during which the spectra are collected. The emission from the sample is coupled to a Princeton Instrument Acton Spectra Pro SP2300 spectrometer equipped with a $LN_2$ cooled Pylon 400 CCD detector.

The external (photocurrent) quantum efficiency (EQE) was determined with a home-built setup comprising an Oriel Halogen lamp, a light chopper coupled to an Oriel monochromator, a Y-fiber adapter directing light to the solar cell and a S2281 Hamamatsu Si reference photodetector. Signals are recorded by two lock-in amplifiers (Signal Recovery 7265, Stanford Research Systems SR830).

Spin-sensitive measurements were performed in a modified X-Band spectrometer (Bruker E300). In PLDMR configuration, samples were illuminated with a glass fiber connected to a cw 532 nm laser. The PLDMR measurements were done at 5 K, provided by a continuous flow helium cryostat (Oxford ESR 900). The PL was detected by a silicon photodiode placed in front of a microwave cavity with optical access (ER4104OR). The change of PL or photocurrent was detected via a Lock-In-Amplifier (Signal Recovery 7230) with the on-off modulated microwave (Anritsu 3694C) as reference. In comparison to the PLDMR setup, in EDMR an electrical signal from the solar cell under test is detected. With a source measuring unit (Keithley 237), every desired point of the *J-V*-characteristic can be chosen to detect the current flowing through the OSC under applied magnetic field. The EDMR measurements were performed under illumination of a white light LED with the approximate equivalent of 1 sun intensity.

3. Results and Discussion

    3.1 Electrical Characterization

*J-V* characteristics of the best devices are presented in Figure 2 a) and solar cell parameters are summarized in Table 1. The averaged parameters of devices, fabricated in the same conditions, are shown in parenthesis. Details of the device optimization are presented in the Supporting Information. The highest performance was obtained for conventional OSCs with a PCE of 9.8% and an external quantum efficiency (EQE) of up to 73% (Figure 2b). Inverted structures did not outperform conventional solar cells with a maximum PCE of 7.5% for freshly prepared inverted devices. During the first days of storage in the nitrogen glovebox, these devices showed a slight increase in $V_{OC}$ and $J_{SC}$ to comparable or better values than the conventional OSCs. The highest PCE for these aged devices was about 8.5%, limited by the lower fill factor (FF) of just 56% in comparison to 66% for conventional devices. The difference in FF may be caused by a non-optimal AZO layer conductivity in comparison to a well-adjusted PEDOT:PSS layer. While the conventional structure featured higher PCE, it showed lower overall stability in comparison to inverted OSCs. The conventional structure retained only 0.2% of efficiency after 20 days of dark storage in nitrogen atmosphere, while the inverted structure declined to 87% of the initial PCE within 100 days. Therefore, further results are mostly obtained for the more stable inverted structure. According to literature, the improvement in performance and stability of the inverted devices can be caused by a vertical phase separation in the active layer[15,16]. The acceptor tends to slowly diffuse to the bottom electrode and in the conventional structure this prevents hole extraction at the cathode. In the inverted structure however, the formation of an ITIC-rich bottom layer is favorable for charge extraction and can lead to both, a slight increase in PCE and enhanced long-term stability. In addition, it has been also shown that the hole transport layer PEDOT:PSS, used in conventional structures. has hygroscopic and acidic nature, and can thus result in a reduction of device stability. The aluminum top contact in the conventional structures can also lead to enhanced OSCs degradation in comparison to air-stable Ag-contacts, utilized in inverted OSCs[13].

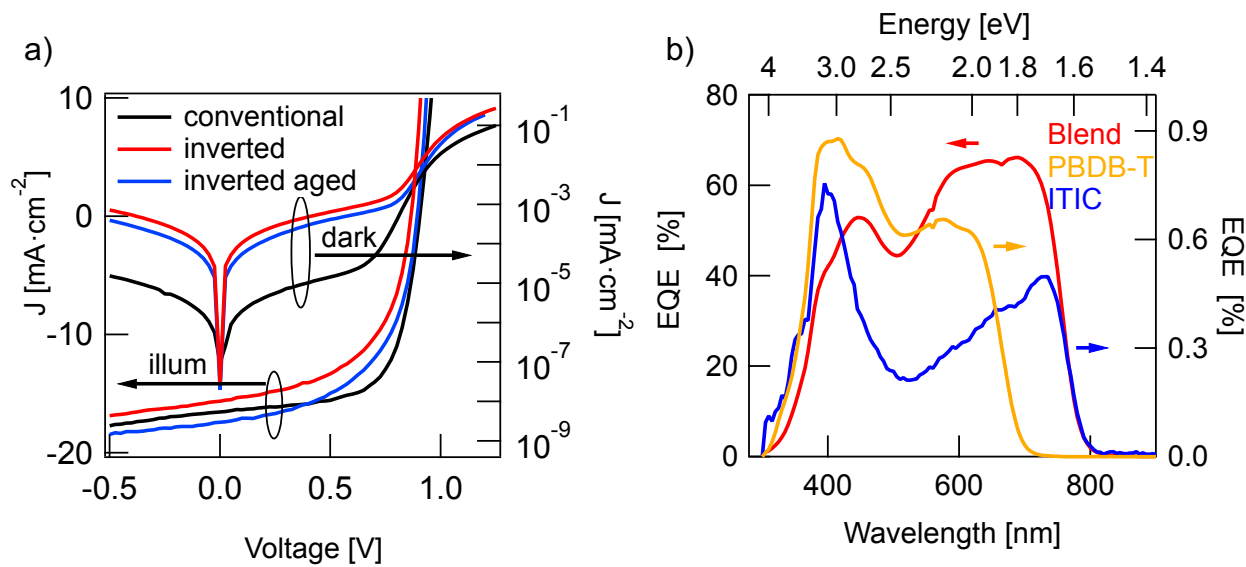

**Figure 2.** Solar cell characteristics. **a)** *J-V*-curves of PBDB-T:ITIC solar cells in dark and under illumination: conventional (black traces), inverted (red) and aged inverted (blue). **b)** EQE spectra for inverted solar cells (red trace) together with diodes from pure PBDB-T (yellow) and pure ITIC (blue).

| structure | (n=) | $V_{OC}$ [mV] | FF [%] | $J_{SC}$ [mA cm$^{-2}$] | PCE [%] |
|---|---|---|---|---|---|
| conventional | 5 | 918 (918) | 66.0 (65.0) | 16.2 (14.7) | 9.8 (8.9±0.9) |
| inverted | 16 | 868 (868) | 53.0 (52.4) | 16.4 (15.3) | 7.5 (7.0±0.3) |
| aged inverted | 11 | 895 (894) | 54.0 (53.6) | 17.4 (16.5) | 8.5 (8.0±0.3) |

**Table 1.** Conventional and inverted solar cell parameters of best devices and averaged parameter values for n=5-16 devices in parenthesis.

### 3.2 Energy Level Determination

Charge generation in OSCs undergoes first, optical absorption into donor and acceptor excited singlet states $S_D$ / $S_A$, followed by interfacial charge transfer to an intermediate CT state. There are several possibilities to assess singlet and CT energies based on rule-of-thumb estimates, direct and indirect measurements that are presented and evaluated below.

One possible method to determine the energy of singlet excited states in pure materials is to determine the midpoint between the low energy peak of the absorption (or EQE) spectra and the high energy peak of the PL spectra. The low energy peaks of the EQE in Figure 2b can be estimated at 1.7 eV for ITIC and 1.96 eV for PBDB-T. The PL spectra in Figure 3a deliver the high energy peaks at 1.6 eV for ITIC and 1.8 eV for PBDB-T. The midpoints are thus 1.65 eV for ITIC and 1.88 eV for PBDB-T.

Since the low energy EQE peaks of these materials are rather undefined (especially in the case of PBDB-T), the above method renders some uncertainty. We thus chose to further corroborate the singlet energy estimations by measuring also the midpoints of the PL and EQE onsets on a logarithmic scale. In Figure 3b the EQE from Figure 2b and the PL from Figure 3a are represented logarithmically over the energy in eV to facilitate pinpointing the onsets. Intensities are scaled as reduced PL / $E$ and reduced EQE × $E$ (see Supporting Information). The EQE (PL) onsets can be read off to be at 770 nm / 1.6 eV (590 nm / 2.1 eV) and 820 nm / 1.5 eV (685 nm / 1.8 eV) for PBDB-T and ITIC, respectively. We note here, however, that the onset of the PL will blue-shift for stronger excitation intensity. Likewise, will the EQE onset redshift if a more sensitive detector would be used. In other words, the external conditions can have a direct effect on these assigned midpoints. However, by comparing both approaches, we can provide decent estimates of the singlet excited state energies as 1.85 eV for the PBDB-T $S_D$ and 1.65 eV for the ITIC $S_A$ with about ±0.05 eV uncertainty.

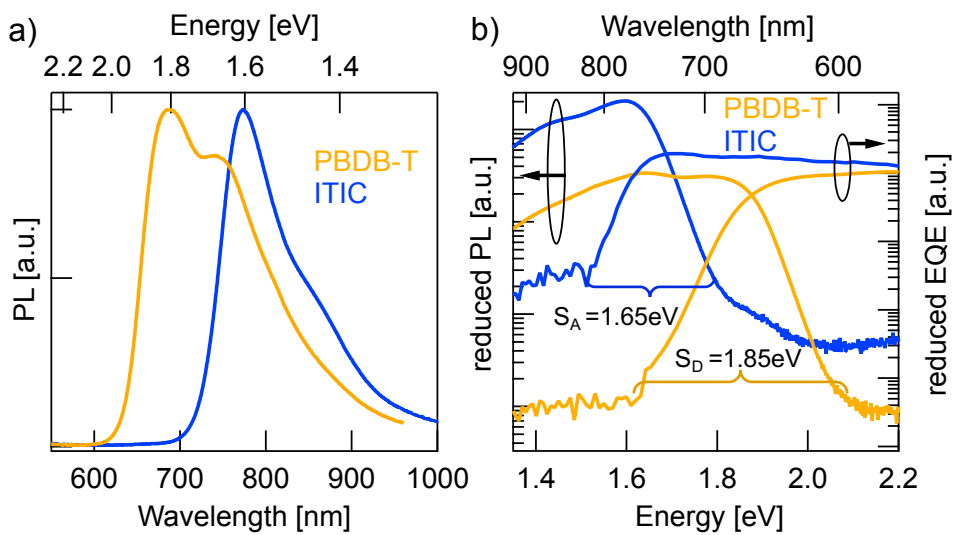

**Figure 3.** Determination of the singlet excited state energies of pure materials. **a)** PL spectra of pure material films: PBDB-T (yellow) and pure ITIC (blue). **b)** The same PL data represented as reduced PL / $E$ together with reduced EQE × $E$ of diodes of the pure materials in logarithmic representation. The onsets of PL and EQE are marked, as well as their midpoint as an estimate for the singlet energy.

The first option for an estimation of the CT energy of the blends is the well-known trend in which the CT energy is about 0.5–0.6 eV higher than $qV_{OC}$ at ambient temperature [17]. With an open-circuit voltage at around 0.9 eV, the CT energy is thus estimated to be 1.45±0.05 eV.

A second approach is to use temperature dependent $J$-$V$ measurements to determine the CT energy. As it has been shown, the extrapolated $qV_{OC}$ value at temperature $T = 0$ K equals $E_{CT}$[18]. We performed $V_{OC}(T)$ measurements for a temperature range from 200 K to 300 K (Figure S3). The extrapolated 0 K value equals 1.40 eV ± 0.05 eV.

The third approach we used to estimate the $E_{CT}$ value employs Marcus theory of the mirror image relationship between optical absorption and emission spectra from and into the CT state as described in detail by Vandewal *et. al*[17] (also see Supporting Information). The midpoint energy of these two spectra has been suggested[19] to embody the most suitable method to determine $E_{CT}$. If the measured spectra are of truly Gaussian shape, then additionally, the reorganization energy λ can be directly deduced from the linewidth of the fitted absorption and emission bands.

Figure 4a shows reduced EQE × $E$ and EL / $E$ spectra for a solar cell and an ITIC diode in inverted device architecture. (EL in linear representation in Figure S2 of the Supporting Information) The solar cell EQE onset is further red-shifted to 900-1000 nm in comparison to the pure material EQEs in Figure 3b. The broader, long-wavelength photocurrent response at energies below the absorbance of pure donor or acceptor films can be ascribed to the direct absorption to CT states. Conversely, with forward biasing, the photovoltaic device is operated as a light-emitting diode with the EL originating from radiative charge recombination via singlet excitons or CT states. The solar cell EL in Figure 4a reaches up to 2 eV (600 nm) without a clear onset and furthermore, it does not have a typical Gaussian shape, but shows a pronounced local bump at around 1.6 eV. The latter can probably be assigned not to CT emission but rather to a superimposed contribution from ITIC singlet exciton EL, which has an emission peak at the same energy. This pronounced emission from ITIC complicates the analysis of EL spectra including reasonable fitting with Marcus theory[20]. To directly address the underlying contribution from CT emission, the ITIC EL spectrum was subtracted from the solar cell EL. For this, the ITIC EL was normalized such that the ITIC EL peak value was corresponding to the local peak value of the solar cell EL. The resulting "subtracted" CT EL signal was used for further analysis in Figure 4b.

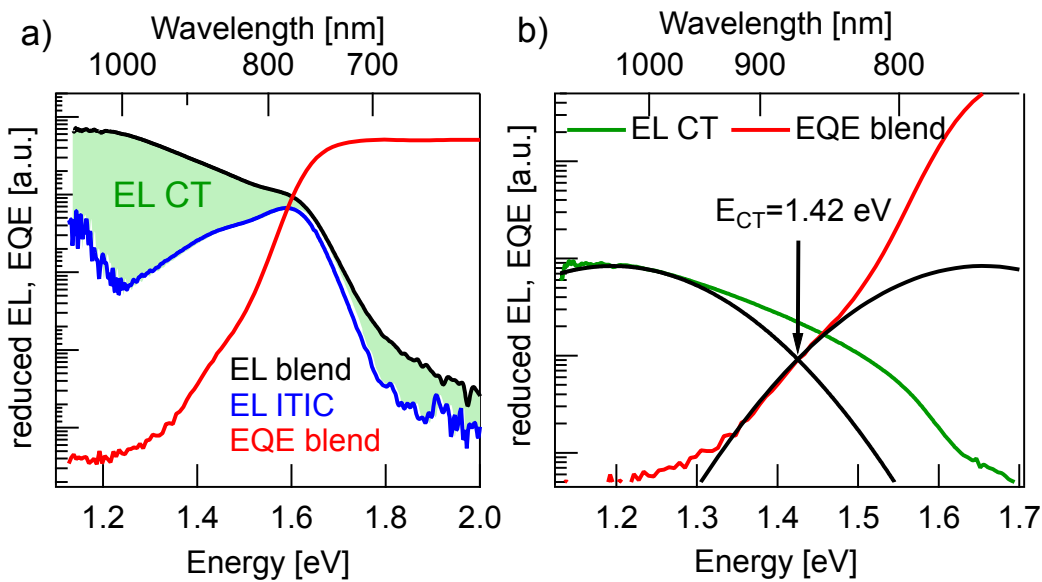

**Figure 4.** Determination of CT state energy. **a)** Reduced EQE × $E$ spectrum for an inverted BHJ solar cell (red) together with EL / $E$ spectra for the same solar cell (black trace) and an ITIC diode (blue). The green shaded area represents the CT EL emission. **b)** Subtracted solar cell EL / $E$ spectrum (green) and reduced EQE × $E$ spectrum (red) together with gaussian fitting curves (black) to determine $E_{CT}$.

In the estimation of $E_{CT}$, gaussian fits to the reduced EQE × $E$ and EL / $E$ spectra in the CT energy regime were performed (Figure 4b). In order to get proper fits, the reorganization energy λ was limited to physically reasonable values in between 0.2 eV to 0.35 eV[17]. For these conditions the best fit in Figure 4b) yields $E_{CT}$ =1.42 eV and λ = 0.23 eV. Due to a very small spectral range where CT absorption dominates the EQE signal (<1.3 eV) and a not perfectly Gaussian form of EL spectra, even after subtraction of the ITIC contribution, this method can (in this case) only give a rough estimate of the value of $E_{CT}$.

In conclusion, all three methods for $E_{CT}$ determination yield values in agreement with their mean value of 1.45 ± 0.05 eV.

### 3.3 Singlet-Triplet Energy Gap by DFT Modelling

We calculated the absorption spectra for a pure PBDB-T tetramer and for ITIC via time-dependent DFT (TD-DFT) within the Tamm-Dancoff approximation (TDA)[21] and employing a polarizable continuum model (PCM)[22] with a dielectric constant of ε = 4.5, in order to take into account the electronic polarization and the solid-state environment. For such calculations, we used the LC-ωhPBE [23] functional in order to resort to a screened RSH (SRSH) functional[24,25] in combination with PCM (See details in Supporting Information). For the polymer donor, we find the first singlet excited state $S_D$ to lie at 2.24 eV above the ground state. The first triplet state $T_D$ is located at 1.84 eV, yielding a singlet-triplet gap of ~0.4 eV. Regarding the ITIC acceptor, its singlet energy $S_A$ falls at 1.95 eV, while $T_{A1}$ is found to be at 1.49 eV (giving rise to a similar singlet-triplet energy splitting of ~0.46 eV). Further, higher-lying ITIC triplet states can be found at 1.69 eV and 2.26 eV. Despite the level of theory used, the calculated singlet excitation energies are overestimated with respect to experiment, by 0.39 eV for the donor PBDB-T and by 0.30 eV for the NFA ITIC molecular acceptor. Part of this discrepancy could be due to solid-state effects and nuclear reorganization effects not included in the modeling. In any case, the results of the calculations combined with the measured singlet energies help drawing a complete Jablonski diagram.

### 3.4 Jablonski Diagram

According to the energy level determination discussed above, the Jablonski diagram of the PBDB-T:ITIC blend can be drawn as shown in Figure 5. The singlet state energy $S_D$ of the donor PBDB-T has the highest value of 1.85 eV and the acceptor ITIC singlet state $S_A$ is at 1.65 eV. Singlet $CT^1$ and triplet $CT^3$ states have degenerate energy values at 1.45±0.05 eV due to the large electron-hole separation distance[26] and oscillation between these two states can occur via intersystem crossing (ISC), e.g. due to hyperfine interaction or slightly different g-factors for electrons and holes. The intended solar cell processes are: optical excitation of donor or acceptor, followed by charge transfer (CT) to the singlet $CT^1$ state, potentially thermally-activated charge separation into free charges and finally charge extraction as photocurrent. Competing loss mechanisms include PL from $S_A$, $S_D$ and $CT^1$, as well as triplet-related loss mechanisms.

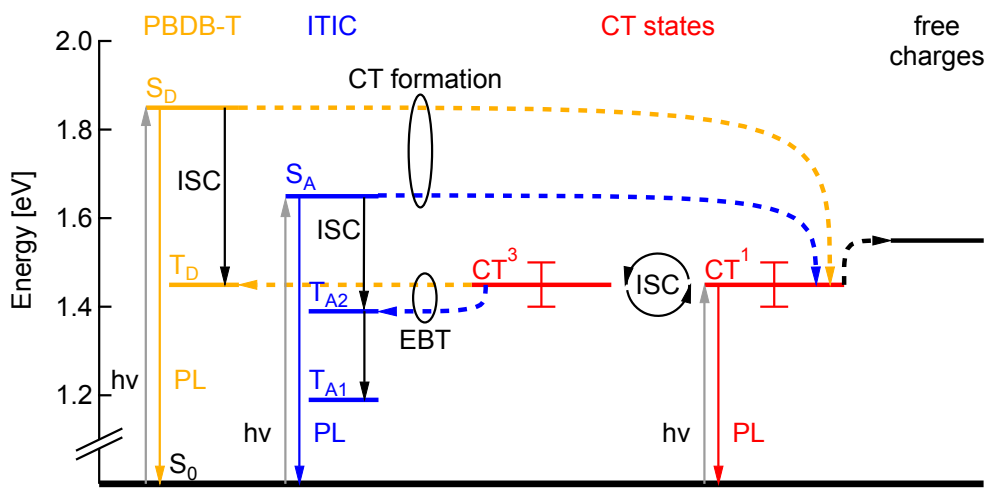

**Figure 5.** Jablonski diagram of the PBDB-T:ITIC blend. $S_D$, $T_D$ - singlet and triplet excited states of the donor PBDB-T (yellow). $S_A$, $T_{A1}$, $T_{A2}$ - singlet and two triplet states of the acceptor ITIC (blue). $CT^1$, $CT^3$ - singlet and triplet charge transfer states (red), ISC - intersystem crossing, EBT - electron back transfer.

Following the simulations described above, $T_D$ lies at 1.45 eV, $T_{A1}$ at 1.19 eV and $T_{A2}$ at 1.39 eV. Triplet states of donor and acceptor may thus be populated either by ISC from the singlet excited states, but also via electron back transfer (EBT) from the triplet $CT^3$, in case this transition is energetically favorable. Consequently, depending on local CT energetics and the efficiency of charge separation, there could be higher or lower population rate of molecular triplet excitons. An efficient triplet state population would lead to additional efficiency losses but would also cause a substantial degradation mechanism in the solar cells, as previously mentioned. To study TE formation we used magnetic resonance methods, that will be discussed in the following part.

### 3.5 Optical Detection of Spin States in Pure Materials and Blends

Optical spectroscopy of triplet excitons alone is not always fully conclusive and cannot readily be applied to fully processed devices. Alternatively, one can take advantage of the paramagnetic properties of TEs [8]. A magnetic field can be used to lift the degeneracy of the three triplet Zeeman sublevels. By applying a microwave field that is resonant with the Zeeman splitting, Zeeman sublevel transitions are induced within the TE manifold. This can in turn modify the overall triplet-triplet annihilation rate, triplet-polaron annihilation rate, triplet relaxation rate, or intersystem crossing rate[9,27,28]. Increasing any of these rates, drives the system of reactions forward, resulting in a change of steady state photoluminescence yield and charge carrier recombination rate. As a result, a change in optical emission or in the solar cells' $J_{SC}$ or $V_{OC}$ is observed and can be used to determine whether TEs are present or not. In this work we used electrically detected magnetic resonance (EDMR) on fully processed solar cells under operating conditions and photoluminescence detected magnetic resonance (PLDMR) on pure material and BHJ films. In comparison to EPR spectroscopy, PLDMR has a higher sensitivity due to the much easier detection of photons in the visible range than in the microwave regime. It is also possible to study the excited states (e.g. triplet excitons) and their recombination processes, such as triplet–triplet annihilation or triplet–polaron annihilation, which usually cannot be probed by conventional EPR. EDMR additionally allows to establish the connection between triplet or other spin states and the photocurrent or photovoltage in real device under working operation conditions, which is not possible by other methods.

The Zeeman splitting of triplets in a magnetic field is shown in Figure 6a. There can be a symmetrical (Fig. 6a, top) or asymmetrical (Fig. 6a, center) splitting. Symmetrical splitting is observed for loosely interacting spins, such as for CT excitons. Asymmetrical splitting is valid for molecular TEs – i.e. in the case of two spins in close vicinity. The dipolar interaction between closely interacting spins leads to an energetic splitting of the spin sublevels even if no external magnetic field is applied[29,30]. Asymmetrical splitting of TE states leads to transitions at two different magnetic field values and resulting spectra will have (at least) two peaks (black trace at Fig. 6a, bottom). Symmetrical splitting of TE states in a magnetic field results in only one transition and thus one peak (Fig. 6a, grey curve, bottom). If both, distant and non-distant spins, are simultaneously present, then the resulting spectrum will be the superposition of a sharp central peak and a broader background signal (Fig. 6a bottom, red envelope). Broader spectral components will therefore be referred to as molecular TE signal and the central narrow peak as $CT^3$ signal.

Additional measurements at lower magnetic fields can be helpful to understand superimposed spectra. Strong dipolar interaction between two spins causes the first order forbidden $\Delta m_s = \pm 2$ transition between the $m_s = -1$ and $m_s = +1$ levels to become slightly allowed. This transition occurs at half the magnetic field required for the allowed transitions, and hence it is called the half-field (HF) transition[31].

The PLDMR analysis performed at 5 K on PBDB-T and ITIC films discloses the presence of different paramagnetic species: molecular TEs ($T_D$ and $T_A$) and $CT^3$ triplet states (Fig. 6b). Both materials also have an almost identical half-field signal at 166 mT.

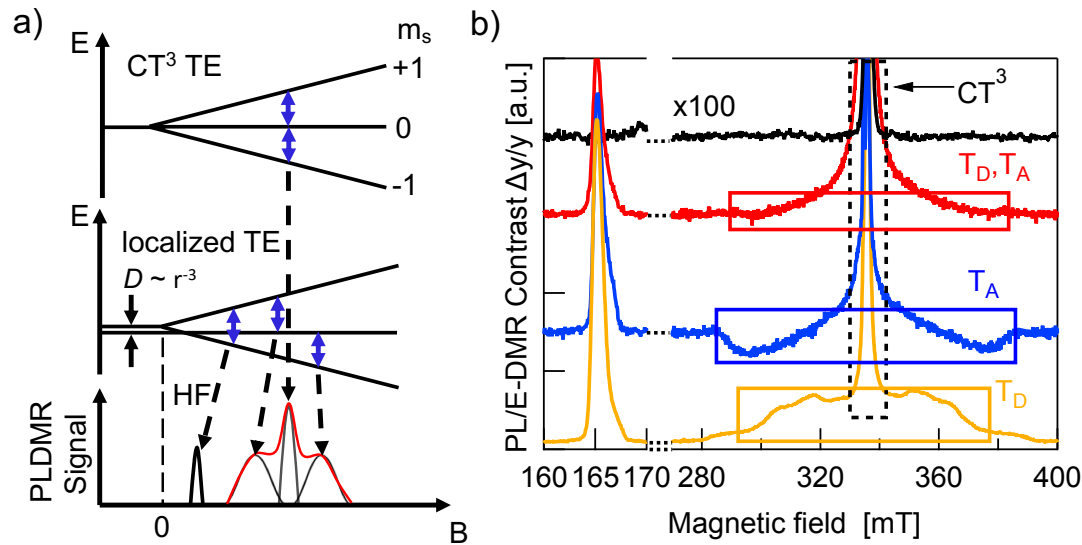

**Figure 6.** Spin-sensitive photoluminescence (PLDMR) and photocurrent (EDMR) spectroscopy of triplet excitons. **a)** Zeeman diagram and zero-field splitting D=0 (top), D>0 (middle) for a triplet S=1 state. Blue arrows indicate possible spin-flip transitions. Red enveloping curve (bottom) symbolises the PLDMR signal shape. HF – (forbidden) half-field transition. **b)** PLDMR contrast $\Delta PL/PL$ of PBDB-T (yellow), ITIC (blue), PBDB-T:ITIC 1:1 blend (red) at 5K; EDMR contrast $\Delta J_{SC}/J_{SC}$ multiplied by factor 100 of inverted

PBDB-T:ITIC SC at 250 K (black). Note, for visibility we cut the CT peaks (dashed box at 340 mT), which is approximately two orders of magnitude higher.

The PLDMR spectrum of the PBDB-T:ITIC BHJ shows a strongly quenched molecular TE contribution and an enhanced CT$^3$ signal. The amplitude of the half-field signal in the blend is also lower than that of pure donor or acceptor. As both pure materials are not easily distinguishable in half-field spectra and molecular TEs of the blend show diminished intensity and cannot clearly be assigned to one or the other material, it is plausible to infer that the formation of molecular TEs occurs on both ITIC and PBDB-T. This fits well with the energetics as presented in Figure 5, as energies of donor and acceptor triplet states are very close to $E_{CT}$. Population of molecular triplet states can occur either via ISC or via EBT from CT to triplet states. The latter being favorable if the energy of the CT state is higher than T$_D$ and T$_{A1}$.

### 3.6 Computation of ITIC Spin Properties

In order to support PLDMR results, spin properties, including zero-field splitting (ZFS) parameters, were calculated at the DFT level of theory on an isolated ITIC molecule in a triplet ground-state (See details in Supporting Information). We assign the broad spectral feature in the ITIC PLDMR spectrum to this intramolecular localized triplet exciton (Figure 7) with an averaged spin-up (in green) – spin-down (in yellow) distance of 1.4-1.5Å.

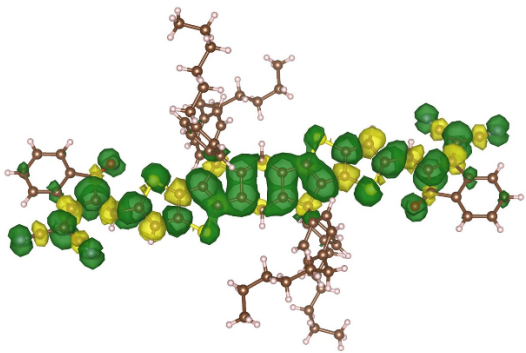

**Figure 7:** Spin density distribution in an ITIC triplet ground state molecule. The green density surface represents the spin-up distribution, while the yellow density surface the spin-down one.

We anticipated that the sharp signal on the ITIC PLDMR spectrum could arise from an intermolecular delocalized triplet exciton state. To assess the nature of the triplet electronic excitations in an ITIC dimer, we performed TD-DFT calculations, taking advantage of the spatial overlap metric $\Phi_S$ between hole and electron densities[32,33]. Pure CT excitations correspond to non-overlapping hole and electron density ($\Phi_S = 0$), while fully localized Frenkel excitations instead lead to $\Phi_S = 1$. By looking at the lowest six triplet excited states of the dimer, we found that each is two-fold degenerate, with the first four transitions being Frenkel excitations that perfectly match those found in the ITIC monomer, i.e. at 1.49 and 1.69 eV. The fifth triplet excited state located at 1.85 eV above the ground state has a rather strong CT character, with the hole density primarily confined on one molecule and the electron density on the other (Figure 8). For this state, we were also able to estimate the e-h capture radius, which can be taken as a proxy for the spin-up – spin-down distance. Such analysis yields an e-h radius of 4.5Å. Thus, we can conclude that intermolecular CT excitations in the pure ITIC phase could be responsible for the sharp peak on the PLDMR spectrum.

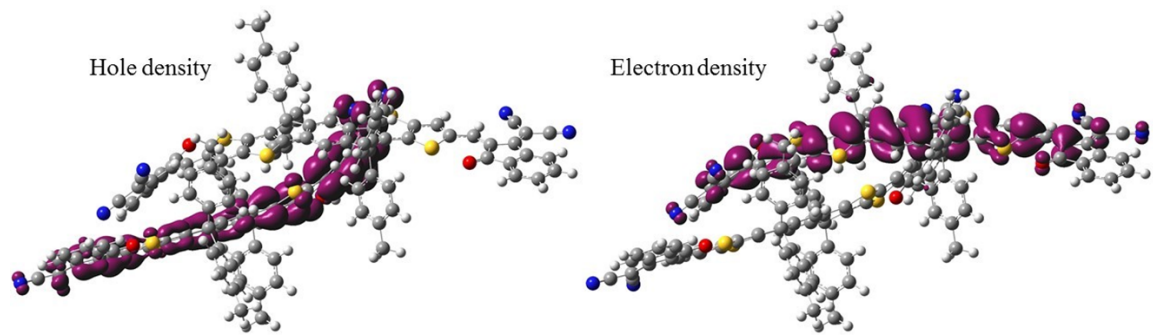

**Figure 8:** Hole and electron density distribution relative to the triplet $T_5$ excited state in an ITIC dimer which shows a strong CT character.

### 3.7 Electrical Detection of Spin States in Solar Cells

After PLDMR has demonstrated the generation of triplet excitons in pure materials and mixed films at low temperature, it is now appropriate to verify triplets also in a solar cell under operating conditions. In this case a direct influence of the triplet excited states on the photocurrent or photovoltage is of high relevance. Therefore, we applied electrically detected magnetic resonance (EDMR) to solar cells in the following. The top-most trace in Figure 6b (black) represents an EDMR spectrum of an OSC in inverted structure, fabricated as described in section 2.1. The measurement was done under white LED illumination at an intensity that yields approximately $J_{SC}$ at 1 sun AM 1.5G illumination ($T = 250$ K). Remarkably, the only observable species in the EDMR spectrum is an intense narrow CT peak. The triplet signal, even the transition at half-field (HF in Fig. 2a), is completely absent. This conclusively shows that molecular triplet excitons, which are generated in mixed films at low temperature, are completely absent in solar cells under operating conditions.

### 3.8 Discussion

These experimental and theoretical results presented above suggest the following scenario: after light absorption, a singlet exciton is generated. In the neat ITIC and PBDB-T films, this singlet exciton can either recombine, undergo ISC to the low-lying triplet state or mostly unlikely dissociate into free charges via a CT state. In the neat films, charge transfer is hindered because of the absence of a suitable electron donor or acceptor. Nevertheless, a substantial CT signal is still observed in PLDMR. This is in line with the low, but detectable EQE and PCE in OSCs based on the neat materials. Conversely, the population of CT states is clearly superior in the blend, suggesting that the charge transfer mechanism is more efficient than ISC. This explains the decrease of the molecular TE signal in favor of an enhanced CT peak in blends. All in all, ISC is still likely to occur at low temperatures because the mobility of singlet excitons is low and this affects the probability of reaching suitable A/D interfaces where CT can occur. On the other hand, if the charge separation of CT states into free charges is slowed down at low temperatures, EBT to molecular TEs is certainly a competitive recombination mechanism.

In devices operated at (or near) ambient temperatures, the photophysical processes can be different: singlet excitons diffuse towards the D/A interface and form CT states more efficiently due to the increased mobility at ambient temperatures. In this case, ISC and EBT will be outperformed by charge separation and extraction. However, there is also the possibility of an increased rate of (non-)radiative recombination before reaching an interface, which is more likely than a slow ISC.

4. Conclusion

In this work we investigated organic solar cells based on the donor PBDB-T and the non-fullerene acceptor ITIC in two different architectures. In the standard architecture, solar cells with structure glass/ITO/PEDOT:PSS/active layer/Ca/Al reached an efficiency of 9.8 % without any additives to the active layer solution. However, they only exhibited a short lifetime of several days. Solar cells in inverted architecture structured as glass/ITO/AZO/active layer/$MoO_3$/Ag showed an improved lifetime and the efficiency retained stable values for at least 100 days. Singlet and triplet state energies of PBDB-T, ITIC and the interfacial CT states were determined by rigorous analysis of photoluminescence, electroluminescence and external quantum efficiency spectra in combination with DFT calculations. Spin properties of ITIC molecules were calculated via DFT modelling, CT and molecular triplets have been assigned with zero-field splitting *D* values and charge distribution. A comprehensive energetic model describing the photophysical processes in PBDB-T:ITIC solar cells was derived. According to a proposed Jablonski diagram, triplet states of donor and acceptor can be populated in the blends either via intersystem crossing from singlet excitons or alternatively via electron back transfer from triplet CT states. We applied spin-sensitive PL measurements to probe the population of triplet states in pure materials and in the blend at low temperatures. The technique allows to distinguish between localized TE and delocalized CT states. Although spin-sensitive PL detection indeed shows a low intensity signal of molecular TEs together with an expected pronounced CT peak in the blends, no molecular TE signals at all could be detected by electrically detected magnetic resonance in solar cells under operating conditions. We attribute this important finding to suppressed TE formation in solar cells, probably due to more efficient charge extraction than the recombination of CT interface states into triplet excitons. These results are also consistent with the high efficiency and stability of solar cells based on PBDB-T:ITIC.

With the highly sensitive tool at hand to probe the populations of CT states and triplet excitons in donor:acceptor absorber blends, we can indeed forecast additional recombination losses due to electron back transfer from CT to localized triplet states. In real devices, the impact of such a process depends on the interaction between electron back transfer and charge carrier extraction, which reflects the electrical properties of the blends and electrode interfaces rather than the donor:acceptor photophysics. Similarly, the active layer degradation can be accelerated by triplet excitons formed, but again in devices it will depend on the measurement's conditions (short-circuit or open-circuit). According to our gained understanding, it is essential to conduct comparative studies on films and devices to clarify the influence of CT and triplet states on the performance of solar cells, provided that we are able to investigate these relevant states selectively and directly.

Acknowledgements

We acknowledge EU H2020 for funding through the Grant SEPOMO (Marie Skłodowska-Curie Grant Agreement 722651). We thank Dr. Andreas Baumann for discussions and help with solar cell processing. Computational resources were provided by the Consortium des Équipements de Calcul Intensif (CÉCI), funded by the Fonds de la Recherche Scientifiques de Belgique (F.R.S.-FNRS) under Grant No. 2.5020.11, as well as the Tier-1 supercomputer of the Fedération Wallonie-Bruxelles, infrastructure funded by the Walloon Region under Grant Agreement No. 1117545. DB is a FNRS Research Director.

# Supporting information for
On the Absence of Triplet Exciton Loss Pathways
in Non-Fullerene Acceptor based Organic Solar Cells

Maria S. Kotova[1], Giacomo Londi[2], Johannes Junker[1], Stefanie Dietz[1], Alberto Privitera[3], Kristofer Tvingstedt[1], David Beljonne[2], Andreas Sperlich[1*] and Vladimir Dyakonov[1]

[1] Experimental Physics 6, Julius Maximilian University of Würzburg, Am Hubland, 97074 Würzburg, Germany

[2] Laboratory for Chemistry of Novel Materials, University of Mons, B-7000 Mons, Belgium

[3] Clarendon Laboratory, Department of Physics, University of Oxford, Oxford OX1 3PU, England, UK

**Device optimization**

Optimization details and corresponding performance parameters of the PBDB-T:ITIC-based OSCs under AM 1.5G illumination (100 mW cm$^{-2}$). Solar cell preparation was tested with and without the additive di-iodooctane (DIO).

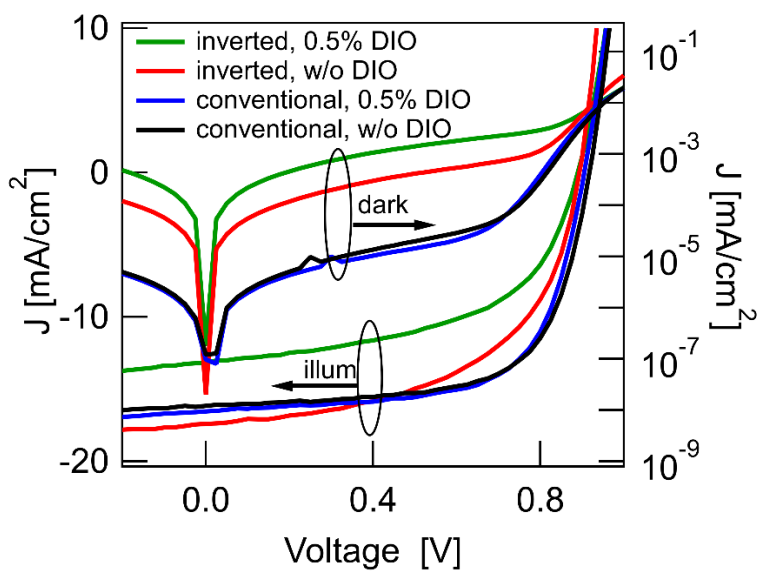

**Figure S1** *J-V*-curves of PBDB-T:ITIC solar cells with and w/o additive DIO in dark and under illumination: inverted with 0.5% DIO (green), inverted w/o DIO (red), conventional with 0.5% DIO (blue), conventional w/o DIO (black).

| structure | (n=) | $V_{OC}$ [mV] | FF [%] | $J_{SC}$ [mA cm$^{-2}$] | PCE [%] |
|---|---|---|---|---|---|
| inverted, 0.5%DIO | 10 | 890 (884) | 54.0 (50.0) | 13.2 (12.8) | 6.3 (5.6±0.4) |
| inverted, w/o DIO | 16 | 868 (868) | 53.0 (52.4) | 16.4 (15.3) | 7.5 (7.0±0.3) |
| aged inverted, w/o DIO | 11 | 895 (894) | 54.0 (53.6) | 17.4 (16.5) | 8.5 (8.0±0.3) |
| conventional, w/o DIO | 5 | 918 (918) | 66.0 (65.0) | 16.2 (14.7) | 9.8 (8.9±0.9) |
| conventional, 0.5% DIO | 10 | 905 (907) | 65.8 (64.6) | 16.5 (15.1) | 9.8 (8.8±0.7) |

**Table S1.** Photovoltaic properties for conventional and inverted PBDB-T:ITIC solar cells with or w/o DIO. Best device values and averaged values for n=5-16 devices in parenthesis.

| AL thickness [nm] | (n=) | $V_{OC}$ [mV] | FF [%] | $J_{SC}$ [mA cm$^{-2}$] | PCE [%] |
|---|---|---|---|---|---|
| 115 | 6 | 918 (916) | 62.7 (60.7) | 16.8 (15.3) | 9.6 (8.5±0.9) |
| 100 | 5 | 918 (919) | 66.2 (65.5) | 16.2 (14.7) | 9.8 (8.9±0.9) |
| 90 | 8 | 918 (916) | 67.8 (66.5) | 12.6 (11.9) | 7.9 (7.3±0.6) |

**Table S2.** Photovoltaic properties of conventional PBDB-T:ITIC solar cells w/o DIO and with different active layer (AL) thickness. Best device values and averaged values for n=5-8 devices in parenthesis.

| Annealing temp. [°C] | (n=) | $V_{OC}$ [mV] | FF [%] | $J_{SC}$ [mA cm$^{-2}$] | PCE [%] |
|---|---|---|---|---|---|
| 70, 30 min | 16 | 868 (868) | 53.0 (52.4) | 16.4 (15.3) | 7.5 (7.0±0.3) |
| 100, 10 min | 8 | 875 (868) | 58.1 (53.8) | 12.6 (12.0) | 6.4 (5.6±0.5) |

**Table S3.** Photovoltaic properties of inverted PBDB-T:ITIC solar cells w/o DIO, but with 10 or 30 minutes of thermal annealing at different temperatures. Averaged values for n=8 or 16 devices in parenthesis.

**Electroluminescence – EL**

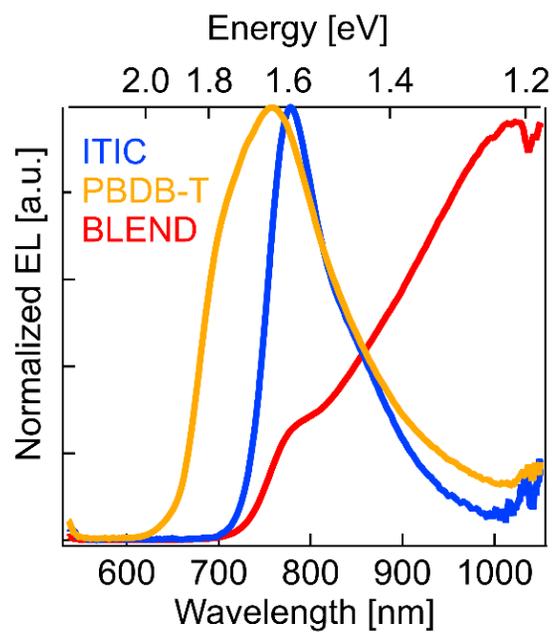

**Figure S2.** Normalized EL spectra of pure PBDB-T (yellow), pure ITIC (blue) and a fully processed solar cell measured as OLED under current injection (red).

**Temperature Dependent $V_{OC}$**

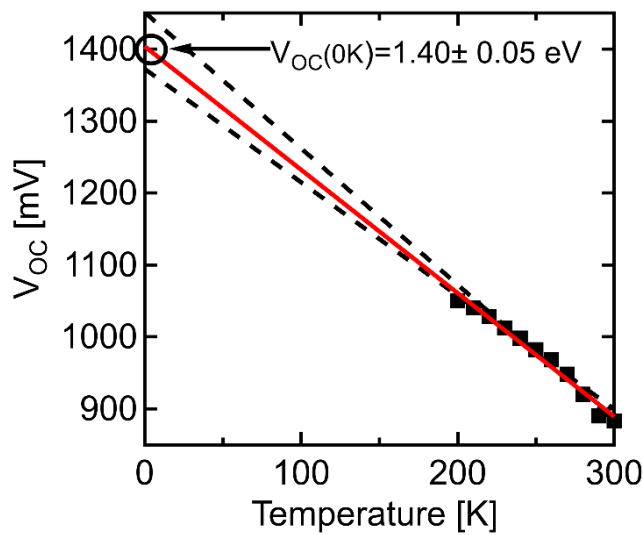

**Figure S3.** Inverted OSC $V_{OC}$ temperature dependence (black dots) and linear fit (red line) with error band (black dashed lines)

**Marcus Theory for Charge Transfer State Absorption and Emission**

In the framework of Marcus theory, the spectral line shape of the charge transfer (CT) state absorption cross section $\sigma(E)$ times photon energy $E$ is described by:

$$\sigma(E)E = \frac{f_\sigma}{\sqrt{4\pi\lambda kT}} \exp\left(\frac{-(E_{CT}+\lambda-E)^2}{4\lambda kT}\right), \qquad (1)$$

where $k$ is Boltzmann's constant and $T$ is the absolute temperature. $E_{CT}$ denotes the free-energy difference between ground state and CTS and $\lambda$ is the reorganization energy, associated with the CT absorption process[1,2].

The counterpart is the CT emission rate $I_f(E)$ per unit energy $E$:

$$\frac{I_f(E)}{E} = \frac{f_{I_f}}{\sqrt{4\pi\lambda kT}} \exp\left(\frac{-(E_{CT}-\lambda-E)^2}{4\lambda kT}\right). \qquad (2)$$

Both $f_{I_f}$ and $f_\sigma$ are not dependent on $E$ and are proportional to the square of the electronic coupling matrix element. The left-hand side of Equations (1) and (2) are called the reduced absorption and emission spectrum respectively and exhibit a mirror image relationship.

**Singlet and Triplet states DFT modelling**

The gas-phase ground-state equilibrium geometry of the donor PBDB-T tetramer and of ITIC were optimized using Density Functional Theory at the range-separated hybrid (RSH) functional level of theory, using the ωB97X-D functional and the 6-31G(d,p) basis set for all the atomic species[3]. In order to speed up the calculations, the alkyl chains in the investigated molecules were replaced with methyl groups. The ITIC structure was taken from the crystallographic data in Ref.4. Excitation energies, oscillator strengths and absorption spectra of the two systems were then investigated by time-dependent DFT (TD-DFT) calculations within the Tamm-Dancoff approximation (TDA) [5] and employing a polarizable continuum model (PCM) [6] with a dielectric constant of ε = 4.5 in order to take the electronic polarization and the solid-state environment into account. For such calculations, we used the LC-ωhPBE [7] functional in order to resort to a screened RSH (SRSH) functional[8,9]. In this approach, ω (the exchange range-separation parameter) was optimally tuned in vacuum and set at 0.1010 Bohr$^{-1}$ for the donor and 0.0970 Bohr$^{-1}$ for the NFA, according to the "gap-tuning" procedure[10,11]. Then, solid-state effects were introduced combining the SRSH functional with PCM and following the relationship $1/\varepsilon = \alpha + \beta$, where $\alpha + \beta$ controls the Hartree-Fock exchange amount at the long-range domain while α quantifies the Hartree-Fock exchange amount at short-range. In our calculations, α was set at 0.2 and, by consequence, β at 0.022. Ground-state optimizations and excitation energies were obtained using the Gaussian16 code[12]. Figure S4 shows the calculated absorption spectra for a pure PBDB-T tetramer and for ITIC.

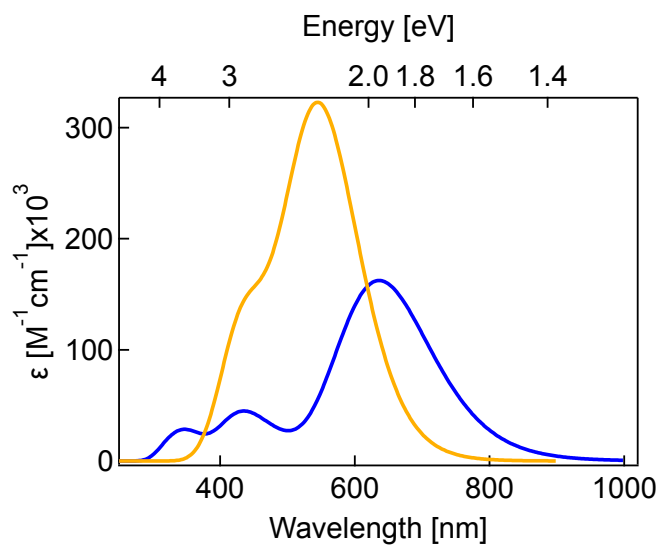

**Figure S4:** TDA-DFT absorption spectra of the donor PBDB-T tetramer (orange) and ITIC (blue). These spectra were obtained with SRSH LC-ωhPBE functional. By mimicking the impact of the solid-state environment, the calculations were carried out within PCM and a dielectric constant ε = 4.5 was set.

In order to corroborate PLDMR results, spin properties, including zero-field splitting (ZFS) parameters, were calculated at the DFT level of theory on an isolated ITIC molecule in a triplet ground-state with the ωB97X-D3 functional and the Def2-TZVP basis set for all the elements[13]. We also took advantage of the RIJCOSX approximation along with the Def2/J and Def2-TZVP/C auxiliary basis set, as implemented in the ORCA 4.2.1 software[14]. In this work, we focused only on the calculation of the direct dipolar spin-spin (SS) contribution to the ZFS tensor; the components of the ***D*** tensor were evaluated using the unrestricted natural orbitals (UNO) obtained from the unrestricted Kohn-Sham orbitals, as suggested in Refs. 15,16. Spin properties are strictly related to the geometries and even small variations of the structural parameters could affect the computed values. For such reason, the ITIC X-ray crystal structure was used for the ZFS calculation, which yielded a $D_{SS}$ of 0.024 cm$^{-1}$ or 25.7 mT and an average spin-up – spin-down distance of 1.4-1.5 Å. As regards the intermolecular delocalized triplet exciton, for this state, we performed an OT-SRSH TDA-DFT/PCM calculation for an ITIC dimer. The dimer was taken directly from the crystallographic data reported in literature[4] and only the hydrogen atoms were optimized at the DFT ωB97X-D/6-31G(d,p) level of theory. Moreover, we were also able to estimate the e-h capture radius of 4.5Å, which can be taken as a proxy for the spin-up – spin-down distance. Since the magnetic dipole-dipole interaction should scale as r$^{-3}$, we estimated the *D* ZFS parameter on the ITIC dimer to be ~1 mT. These findings are in line with the results from the EasySpin calculations (*vide infra*).

**Modeling of Triplets in Pure Materials by EasySpin**

The magnetic resonance spectrum of a triplet can be calculated on the basis of the eigenvalues and eigenvectors obtained from the diagonalization of the spin Hamiltonian that includes the anisotropic Zeeman term and the electron dipolar term, also called zero-field splitting (ZFS) term:

$$H = g\mu_B BS + SDS$$

where ***g*** is the g-tensor related to Zeeman interaction, ***D*** is the dipolar interaction tensor, ***B*** is the magnetic field vector, ***S*** is the spin operator, and $\mu_B$ is the Bohr magneton[17,18]. The eigenvalues *X, Y* and *Z* of the ***D*** tensor are commonly expressed in terms of the ZFS parameters *D* and *E* that are defined as *D* = -3/2Z and *E* = 1/2(Y-X). The *D* parameter is related to the average interaction between the two unpaired spins, and therefore contains information about the mean distance of the two and thus the triplet state delocalization. The *E* parameter describes the off-axial interaction strength for systems with symmetry lower than axial. In general, there are two allowed transitions between the three triplet sublevels ($\Delta m_s = \pm 1$) that correspond to two peaks of the magnetic resonance spectrum. Both, the Zeeman and the dipolar terms in the spin Hamiltonian, however, depend on the relative orientation between the molecules and the magnetic field ***B*** and therefore, in a disordered material like an organic film, the resulting spectrum is the sum of contributions from all randomly-oriented molecules. This is commonly referred to as a powder pattern.

The PLDMR spectra of PBDB-T and ITIC thin films are shown in Figure S5. Both contain a broad spectral feature that can be assigned to localized molecular triplet excitons together with a narrow peak in the center that stems from CT TE and exciton-charge interaction. This contribution is cut and neglected for PBDB-T. The EasySpin[19] spectral simulation of the PLDMR spectrum of PBDB-T is obtained using the following ZFS parameters: *D* = 52.0 mT and *E* = 5.6 mT. The spectral simulation of the ITIC spectrum includes two sets of parameters. The first triplet, corresponding to the broader signal, has the following ZFS parameters: *D* = 45.5 mT and an *E* = 15.0 mT. The narrow peak of the ITIC spectrum contains two contributions: a very narrow peak that, like for PBDB-T, is neglected and a less narrow peak that we assign to a delocalized triplet exciton. The simulation of this ITIC triplet state provides $D = 0.8$ mT and $E = 0$. We attribute these two triplets to a strongly localized molecular triplet exciton (broad spectrum with strong dipolar interaction) and more delocalized TE state with some CT character (narrow spectrum with weak dipolar interaction). Two TE states with slightly different energy and different delocalization are predicted by theory and detected here with magnetic resonance.

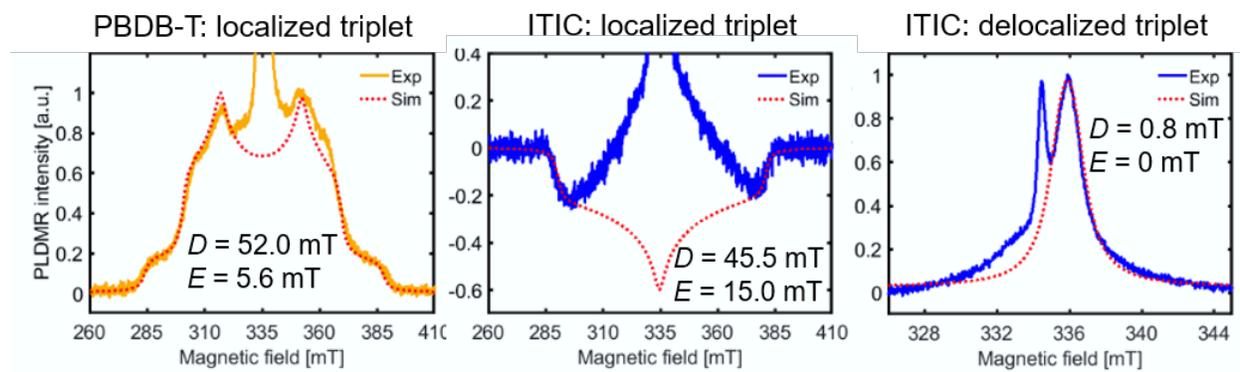

**Figure S5.** PLDMR spectra for PBDB-T (yellow) and ITIC (blue) triplet excitons together with spectral simulations. While the spectrum of PBDB-T and the narrow spectral feature of ITIC can be fitted quite accurately, the sign of the broad spectral features of ITIC cannot be reproduced. The peaks, shoulders and turning points are accurately determined, which is sufficient to extract TE parameters $D$ and $E$. The sign and intensity of a PLDMR spectrum can however be dependent on orientation dependent rate constants and are not easily reproducible by simulation.

## Supporting Information References